\documentclass[aps,prd,twocolumn,superscriptaddress,amsmath,showpacs]{revtex4-1}

\usepackage{graphicx}
\usepackage{dcolumn}
\usepackage{bm}
\usepackage{natbib}
\usepackage{verbatim}
\usepackage{braket}
\usepackage{enumitem}
\usepackage{lipsum}
\usepackage[usenames,dvipsnames]{color}
\usepackage{hyperref}
\hypersetup{
    colorlinks=true, 
    linkcolor=NavyBlue, 
    citecolor=Magenta}
\usepackage{soul,xcolor}
\usepackage{afterpage}

\newcommand{\be}{\begin{equation}}
\newcommand{\ee}{\end{equation}}
\newcommand{\bea}{\begin{eqnarray}}
\newcommand{\eea}{\end{eqnarray}}

\newcommand{\dif}{\text{d}}

\newcommand{\neff}{N_\textup{eff}}
\newcommand{\alens}{A_\textup{L}}
\newcommand{\aeisw}{A_{e\text{ISW}}}
\newcommand{\alisw}{A_{l\text{ISW}}}
\newcommand{\LCDM}{\Lambda\text{CDM}}
\newcommand{\TT}{Planck\,\text{TT+lowP}}
\newcommand{\TTTEEE}{Planck\,\text{TT,TE,EE+lowP}}

\newcommand{\tu}{\textup}

\urldef\mgcamb\url{http://www.sfu.ca/~aha25/MGCAMB.html}

\begin{document}

\title{Constraints on the Early and Late Integrated Sachs-Wolfe effects from
Planck 2015 Cosmic Microwave Background Anisotropies angular power spectra}

\author {Giovanni Cabass}
\affiliation{Physics Department and INFN, Universit\`a di Roma ``La Sapienza'', P.le Aldo Moro 2, 00185, Rome, Italy}

\author {Martina Gerbino}
\affiliation{Physics Department and INFN, Universit\`a di Roma ``La Sapienza'', P.le Aldo Moro 2, 00185, Rome, Italy}

\author {Elena Giusarma}
\affiliation{Physics Department and INFN, Universit\`a di Roma ``La Sapienza'', P.le Aldo Moro 2, 00185, Rome, Italy}

\author {Alessandro Melchiorri}
\affiliation{Physics Department and INFN, Universit\`a di Roma ``La Sapienza'', P.le Aldo Moro 2, 00185, Rome, Italy}

\author {Luca Pagano}
\affiliation{Physics Department and INFN, Universit\`a di Roma ``La Sapienza'', P.le Aldo Moro 2, 00185, Rome, Italy}

\author {Laura Salvati}
\affiliation{Physics Department and INFN, Universit\`a di Roma ``La Sapienza'', P.le Aldo Moro 2, 00185, Rome, Italy}

\begin{abstract}
\noindent The Integrated Sachs-Wolfe (ISW) effect predicts 
additional anisotropies in the Cosmic Microwave Background due to time variation of the gravitational potential when the expansion of the universe is not matter dominated. The ISW effect is therefore expected in the early universe, due to the presence of relativistic particles at recombination, and in the late universe, when dark energy starts to dominate the expansion. Deviations from the standard picture can be parameterized by $A_{e\text{ISW}}$ and $A_{l\text{ISW}}$, which rescale the overall amplitude of the early and late ISW effects. Analyzing the most recent CMB temperature spectra from the Planck 2015 release, we detect the presence of the early ISW at high significance with $A_{e\text{ISW}} = 1.06\pm0.04$ at 68\% CL
and an upper limit for the late ISW of $A_{l\text{ISW}} < 1.1$ at 95\% CL. The inclusion of the recent polarization data from the Planck experiment \textcolor{black}{results in $A_{e\text{ISW}} = 0.999\pm0.028$ at 68\% CL, in better agreement with the value $A_{e\text{ISW}} = 1$ of a standard cosmology.} 
When considering the recent detections of the late ISW coming from 
correlations between CMB temperature anisotropies and weak lensing, a value of $A_{l\text{ISW}}=0.85\pm0.21$ is predicted at 68\% CL, showing a $4\sigma$ evidence. We discuss the stability of our result in the case of an extra relativistic energy component parametrized by the effective neutrino number $N_\textup{eff}$ and of a CMB lensing amplitude $A_\textup{L}$.
\end{abstract}


\maketitle

\section{Introduction} \label{sec:intro}

\noindent Already in $1966$, only two years after the discovery of the Cosmic Microwave Background (hereafter, CMB) radiation
\cite{penziaswilson},  R. K. Sachs and A. M. Wolfe presented the first computations of the gravitational redshift of
CMB photons by linear matter perturbations \cite{sachswolfe}. This so called ``Sachs-Wolfe'' (SW) effect can be identified
in two regimes: the Non-Integrated SW (NISW) effect and the Integrated SW effect (ISW).
The NISW is the predominant source of fluctuations in the CMB on scales larger than $\sim 10$ degrees. This effect,
measured for the first time by the COBE satellite in $1992$ \cite{COBE}, occurs at the last scattering surface and provides
the first indication for a nearly scale invariant spectrum of primordial fluctuations, as expected in inflationary theory (see e.g. \cite{inflationcobe1, inflationcobe2}).

The ISW, on the contrary, is a ``secondary'' source of CMB fluctuations, always subdominant with respect to primary sources: it is produced
between the last scattering surface and today, and it gives a non-zero contribution only if the expansion of the universe is not entirely driven
by a non-relativistic matter component. Therefore it will be present after CMB decoupling
(produced by 
\textcolor{black}{the non-negligible relativistic energy component in the total energy density} -- early ISW), and at recent
times when the expansion of the Universe starts to be affected by dark energy (late ISW).

Both $e$ISW and $l$ISW provide an excellent probe for ``new physics''. A measurement of a late ISW is indeed an evidence for a 
non-dark matter dominated expansion of the late Universe, confirming the existence of a ``dark energy'' component.
The $l$ISW, combined with other cosmological observables, could also be used to constrain dark energy parameters as
its equation of state or effective sound speed (see e.g. \cite{iswparameters1, iswparameters2, iswparameters3}).
Moreover, the use of the $l$ISW to constrain the neutrino mass has been \mbox{proposed by \cite{iswjulien}}.

The $e$ISW, on the contrary, probes the 
amount of energy stored 
in relativistic degrees of freedom at recombination.
The presence of extra-light particles like sterile neutrinos or thermal axions at such epoch, then, can change its amplitude.
The early ISW can also be used to constrain modified gravity models as discussed, for example, in \cite{Zhang:2005vt}.


The $l$ISW has been detected for the first time in \cite{crittenden}, by cross-correlating the map of the CMB sky measured by the WMAP satellite with number counts of radio galaxies in the NVSS survey and with the hard X-ray background measured by the HEAO-1 satellite.

This detection has then been confirmed several times in the past years by cross-correlations
with different datasets \cite{iswmeasurements1,iswmeasurements2,iswmeasurements3,iswmeasurements4,iswmeasurements5,iswmeasurements6,iswmeasurements7,iswmeasurements8,iswmeasurements9}. The last analysis obtained by the Planck collaboration \cite{planckisw} found a 
$\sim 4\sigma$ indication for $l$ISW, with an amplitude in agreement 
with a cosmological constant making up the entirety of the dark energy component.

The $e$ISW cannot be probed directly, but it affects the CMB angular spectrum of temperature anisotropies (see e.g.
\cite{hou} and the discussion in the next Section). Constraints on the amplitude of the $e$ISW coming from the WMAP satellite 
have been presented in \cite{hou}.

In this paper we present new constraints on the $l$ISW and the $e$ISW effects from the recent measurements of the
CMB temperature and polarization angular power spectrum provided by the Planck satellite
, 
and also discuss degeneracies with other parameters.
Most notably, we found a correlation between the amplitude of the $e$ISW and the effective lensing
parameter $A_\textup{L}$ 
in discrepancy with the standard value at about $\sim 2$ standard deviations.

The paper is organized as follows: in the next Section we 
describe the physics of the ISW effect and the parametrization we have used.
In Section \ref{sec:method} we present our data analysis method, in Section \ref{sec:results} we discuss our results and, finally, in Section \ref{sec:concl} we
derive \mbox{our conclusions.}

\section{The ISW effect} \label{sec:isw-theory}

\noindent The Integrated Sachs-Wolfe (ISW) effect is a contribution to the CMB temperature anisotropy given by the interaction of photons with time-dependent gravitational potentials. At multipole $\ell$ and linear order in temperature perturbations one has that \cite{footnote:1}
\begin{equation}
\label{eq:1}
\Theta_\ell^\textup{ISW}(k) = \int_0^{\eta_0}\dif\eta\,e^{-\tau(\eta)}\big\{\dot{\Psi}(k,\eta) - \dot{\Phi}(k,\eta)\big\}j_\ell(k\Delta\eta)\,\,,
\end{equation}
where $\tau$ is the optical depth, $\eta_0$ is the current conformal time and $\Delta\eta\equiv\eta_0 - \eta$. For times much earlier than recombination ($\eta\ll\eta_\textup{rec}$), CMB photons are tightly coupled to electrons and protons by Compton scattering: this makes $e^{-\tau(\eta)}$ small enough that the ISW effect is negligible.

\subsection{Early ISW -- theory} \label{subsec:eisw-theory}

\noindent Eq.~\eqref{eq:1} shows how there is a non-vanishing ISW effect in presence of time dependent gravitational potentials $\Psi$ and $\Phi$. 
\textcolor{black}{For modes that cross the 
horizon well into matter domination, the gravitational potentials are constant in time. So one expects the ISW to be mainly present at times after recombination (since the energy density of relativistic matter is still considerable at that time).}
Because of this, one can estimate its contribution to multipole $\ell$ by evaluating the Bessel function at $\eta\sim\eta_\textup{rec}$: the result is (approximating $\Phi\approx-\Psi$)

\begin{equation}
\label{eq:1-bis}
\Theta_\ell^{e\text{ISW}}(k)\approx 2j_\ell(k\Delta\eta_\textup{rec})\big\{\Psi(k,\eta_\textup{MD}) - \Psi(k,\eta_\textup{rec})\big\}\,\,, 
\end{equation}
where $\eta_\textup{MD}$ is a time 
late at matter domination. From Eq.~\eqref{eq:1-bis} one can see that \cite{Bowen:2001in}:
\begin{itemize}[leftmargin=*]
\item \textcolor{black}{the early ISW adds in phase with the Sachs-Wolfe primary anisotropy, given by
\begin{equation}
\label{eq:1-bis-bis}
\Theta_\ell^{\text{SW}}(k)=j_\ell(k\Delta\eta_\textup{rec})\big\{\Theta_0(k,\eta_\textup{rec}) + \Psi(k,\eta_\textup{rec})\big\}\,\,.
\end{equation}
We can see this from the fact that both anisotropies are multiplied by the same Bessel function.}
This will increase the height of the first acoustic peaks, with the first one being boosted more than the others. The reason is that at times right after recombination, perturbations with $k\ll 1/\eta_\textup{rec}$ do not evolve, while perturbations with $k\gg 1/\eta_\textup{rec}$ are averaged out when integrated along the photon trajectory. This means that the \textcolor{black}{dominant} contribution to the early ISW effect is due to perturbations with $k\sim 1/\eta_\textup{rec}$, that approximately corresponds to the first acoustic peak;
\item the effect of $\Theta_\ell^{e\text{ISW}}(k)$ on the angular anisotropy $C_\ell$ is suppressed by the factor 
\begin{equation}
\label{eq:1-ter}
\frac{\rho^2_\textup{rad}(\eta_\textup{rec})}{\rho^2_\textup{m}(\eta_\textup{rec})} = \bigg(\frac{1 + z_\textup{rec}}{1 + z_\textup{eq}}\bigg)^2\,\,.
\end{equation}
\textcolor{black}{Therefore, even if neutrinos and other relativistic species decoupled from the primordial plasma earlier than the photons, the ISW will still depend on the number of relativistic degrees of freedom at recombination: an increase of the amount of radiation during this epoch (i.e. an effective number of relativistic species $N_\textup{eff} > 3.046$) will delay the advent of matter domination, make $z_\textup{eq}$ smaller, and result in a larger amplitude of the early ISW effect.}
\end{itemize}

This is one of the main reasons why the Cosmic Microwave Background is sensitive to the redshift of matter-radiation equality (and then to the amount of radiation at recombination), thus opening the possibility of constraining the number of extra relativistic species with CMB experiments.

\subsection{Late ISW -- theory} \label{subsec:lisw-theory}

\noindent\textcolor{black}{The late ISW effect is active at more recent times, when dark energy starts to play a role and the gravitational potentials are decreasing, and its contribution to the CMB power spectrum is sizable at large scales only \cite{footnote:2}. The observable effects of the ISW, in the times dominated by dark energy, are mainly the following \cite{Manzotti:2014kta}:}
\begin{itemize}[leftmargin=*]
\item\textcolor{black}{focusing on scales corresponding to galaxy clusters, where gravitational perturbations start growing, the CMB photons experience an ISW effect caused by the time-dependence of the gravitational potential inside these non-linear structures. Therefore one expects to find a correlation between $C_\ell^\textup{ISW}$ and the density contrast observed by surveys \cite{Corasaniti:2005pq, Cabre:2006qm}. These correlations can be used to distinguish between the standard $\Lambda\text{CDM}$ universe and models that try to explain the present day acceleration through modifications of gravity \cite{Lue:2003ky, DiValentino:2012yg};}
\item\textcolor{black}{the gravitational potentials that redshift CMB photons (late ISW) are the same that cause the weak lensing distortions: the interplay between these two effects gives rise to a non-Gaussian contribution, which is encoded in the lensing-induced bispectrum between small and large angular scales \cite{Hu:2001kj}.}
\end{itemize}
\textcolor{black}{The correlation with these LSS tracers has been investigated in \cite{planckisw}, which studied the cross-correlations of the temperature anisotropies with both lensing potential and galaxy number counts, showing that they yeld a $4\sigma$ detection of the late ISW. More precisely, temperature-lensing correlations result in $A_{l\text{ISW}} = 1.04\pm0.33$, while including galaxy number counts gives $A_{l\text{ISW}} = 1.00\pm0.25$.}


\subsection{Parametrization of early and late ISW effects} \label{subsec:eisw+lisw-parametrization}

\noindent In this paper we consider a parametrization of the ISW amplitude in terms of two parameters $A_{e\text{ISW}}$ and $A_{l\text{ISW}}$, which rescale the contribution at early ($A_{e\text{ISW}}$) and late ($A_{l\text{ISW}}$) times in the following way: we introduce in the integrand of Eq.~\eqref{eq:1} a function $f(\eta)$ given by
\begin{equation}
\label{eq:7}
f(\eta) =
\begin{cases}
A_{e\text{ISW}} &\text{for $z > 30\,\,,$} \\
A_{l\text{ISW}} &\text{for $z < 30\,\,,$}
\end{cases}
\end{equation}
where the standard scenario is given by $A_{e\text{ISW}} = A_{l\text{ISW}} = 1$. The reason why we have chosen $z = 30$ as a turning point between the early and late contributions is merely a phenomenological one: plotting the integrand of Eq.~\eqref{eq:1} as a function of redshift with the \texttt{camb} code \cite{url_of_camb}, one can see that its minimum lies near $z = 30$. 

\section{Data Analysis Method} \label{sec:method}

\noindent We perform a Markov-chain Monte-Carlo (MCMC) analysis, making use of the publicly available code \texttt{cosmomc} \cite{Lewis:2013hha, Lewis:2002ah}. Our baseline model is the standard six-parameter $\Lambda$CDM model, which includes the baryon density $\Omega_b h^2$, the cold dark matter density $\Omega_c h^2$, the sound horizon angular scale $\theta$, the reionization optical depth $\tau$, the amplitude and spectral index of the primordial power spectrum of scalar perturbations $\ln[10^{10} A_s]$ and $n_\textup{s}$. We then include the two amplitudes $A_{e\text{ISW}}$ and $A_{l\text{ISW}}$ of Eq.~\eqref{eq:7}.

We firstly fix one of the two amplitudes to the standard expected value and let the second one to vary freely, but also explore the case of the two amplitudes varying jointly. In addition, we consider other one-parameter extensions to this $\Lambda\text{CDM}+A_\textup{ISW}$ model, by varying separately the gravitational lensing amplitude $A_\textup{L}$ \cite{Calabrese:2008rt}, the primordial helium abundance $Y_P$ (assuming it to be an independent parameter in a non-standard BBN framework) and $T_\textup{CMB}$ (the blackbody temperature of the CMB at the current epoch). When not varied, these parameters are fixed in agreement with the standard cosmological scenario, namely:
\begin{itemize}[leftmargin=*]
\item $A_\textup{L} = 1$;
\item $N_\textup{eff} = 3.046$;
\item $Y_P$ as a function of $\Omega_b h^2$ and the effective number of relativistic species $N_\textup{eff}$ equal to $3.046$ (as expected from the standard BBN);
\item $T_0 = 2.7255\,\text{K}$ \cite{Fixsen:2009}.
\end{itemize}

We impose flat priors, but also check the impact of a gaussian prior
$A_{l\text{ISW}} = 1.00\pm 0.25$ (which will be denoted by the ``prior'' label in the following plots and tables). This prior is consistent with the 68\% CL bounds on the same parameter from \cite{planckisw}, where the ISW-lensing bispectrum induced on the Gaussian CMB anisotropies by
the lensing effect is estimated by
cross-correlating the Planck CMB maps with the Planck map of the lensing potential \cite{planckisw}.

We test the following datasets: the high-$\ell$ Planck temperature and polarization power spectra in the range $30\leq\ell<2500$ (hereafter \textit{Planck} TT and \textit{Planck} TT, TE, EE) combined with the low-$\ell$ Planck temperature and polarization power spectra in the range $2\leq\ell<29$ (denoted as lowP) 
\cite{Aghanim:2015wva}.
Regarding polarization spectra at high $\ell$, we also test the WMAP power spectra in temperature and polarization \cite{Bennett:2013} up to $\ell=1200$
. When $T_0$ is varied, we also add information from baryon acoustic oscillation (BAO) as reported in \cite{Planck:2015xua}, in order to break degeneracies among cosmological parameters.

\section{Results} \label{sec:results}


\subsection{Early ISW -- results} \label{subsec:eisw-results}

\noindent We start from considering the case in which only the early ISW effect is left free to vary. The results of our analysis are shown in Tabs.~\ref{tab:big_tab-2} and \ref{tab:big_tab-2-TTTEEE} in which we report the 68\% CL around the mean value of the posterior. 

\begin{table*}
\caption{Constraints at 68\% CL on the cosmological parameters in the extended 
$\Lambda$CDM model explored here using the \textit{Planck} TT+lowP dataset.}
\begin{center}

\begin{tabular}{c|c|c|c}
\hline \hline
Parameter	  &$\LCDM+\aeisw$   &$\LCDM+\neff+\aeisw$     &$\LCDM+\alens+\aeisw$\\
\hline \hline
\textcolor{black}{$\Omega_\textup{b}h^2$}      &\textcolor{black}{$0.0218\pm0.0004$}             &\textcolor{black}{$0.0218\pm0.0005$}         &\textcolor{black}{$0.0225\pm0.0005$}\\
$\Omega_\textup{c}h^2$      &$0.1201\pm0.0022$             &$0.1204\pm0.0039$             &$0.1170\pm0.0027$\\
$100\theta$                          &$1.04072\pm0.00049$          &$1.04071\pm0.00056$         &$1.04126\pm0.00056$\\
$\tau$                                   &$0.076\pm0.019$                  &$0.077\pm0.022$                 &$0.059\pm0.020$\\
$n_\textup{s}$                      &$0.9724\pm0.0080$              &$0.974\pm0.016$                 &$0.9750\pm0.0081$\\
$\ln[10^{10}A_\textup{s}]$    &$3.080\pm0.037$                  &$3.083\pm0.048$                 &$3.045\pm0.041$\\
$N_\textup{eff}$                    &$\equiv 1$                             &$3.08^{+0.29}_{-0.34}$        &$\equiv 3.046$\\
$A_\textup{L}$                      &$\equiv 1$                             &$\equiv 1$                            &$1.216\pm0.11$\\
$A_{e\text{ISW}}$               &$1.064^{+0.042}_{-0.043}$     &$1.065\pm0.043$                &$1.018\pm0.046$\\
\hline
\hline
\end{tabular}
\end{center}
\label{tab:big_tab-2}
\end{table*}


\begin{table*}
\caption{Constraints at 68\% CL on extensions of the
$\Lambda$CDM model for the \textit{Planck} TT,TE,EE+lowP dataset.}
\begin{center}
\begin{tabular}{c|c|c|c}
\hline \hline
Parameter	  &$\LCDM+\aeisw$    &$\LCDM+\neff+\aeisw$   &$\LCDM+\alens+\aeisw$\\
\hline \hline
\textcolor{black}{$\Omega_\textup{b}h^2$}     &\textcolor{black}{$0.0222\pm0.0002$}                    &\textcolor{black}{$0.0222\pm 0.0003$}     	                   &\textcolor{black}{$0.0225\pm 0.0002$}\\
$\Omega_\textup{c}h^2$     &$0.1199\pm0.0015$                   &$0.1189\pm 0.0031$                                    &$0.1183_{-0.0015}^{+0.0016}$\\
$100\theta$                         &$1.04072\pm0.00031$               &$1.04087\pm 0.00045$	                          &$1.04095\pm 0.00032$\\
$\tau$                                  &$0.081\pm0.017$                       &$0.080\pm 0.018$                                       &$0.056_{-0.020}^{+0.021}$\\
$n_\textup{s}$                     &$0.9638\pm0.0058$                   &$ 0.961\pm 0.010$                                      &$0.967\pm0.0055$\\
$\ln[10^{10}A_\textup{s}]$   &$3.098\pm0.033$                       &$3.091\pm 0.038$                                      &$3.042^{+0.043}_{-0.040}$\\
$N_\textup{eff}$                  &$\equiv 1$                                   &$2.99^{+0.20}_{-0.21}$                              &$\equiv 3.046$\\
$A_\textup{L}$                    &$\equiv 1$                                   &$\equiv 1$                                                   &$1.182_{-0.086}^{+0.076}$\\
$A_{e\text{ISW}}$               &$0.999\pm 0.028$                      &$1.002\pm 0.028$                                       &$0.988\pm0.027$\\
\hline
\hline
\end{tabular}
\end{center}
\label{tab:big_tab-2-TTTEEE}
\end{table*}

\begin{figure*}
\begin{center}
\begin{tabular}{c c}
\includegraphics[width=0.98\columnwidth]{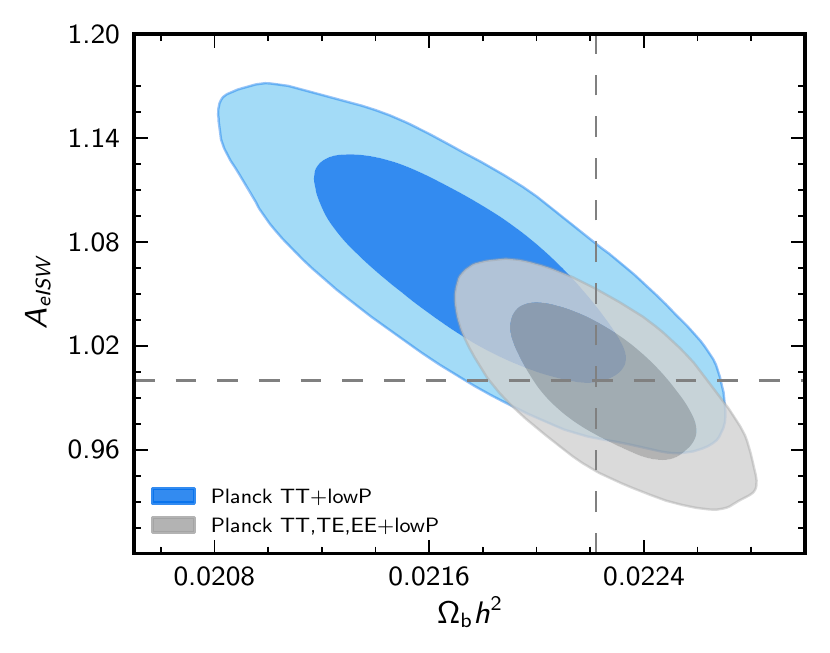}&\includegraphics[width=0.98\columnwidth]{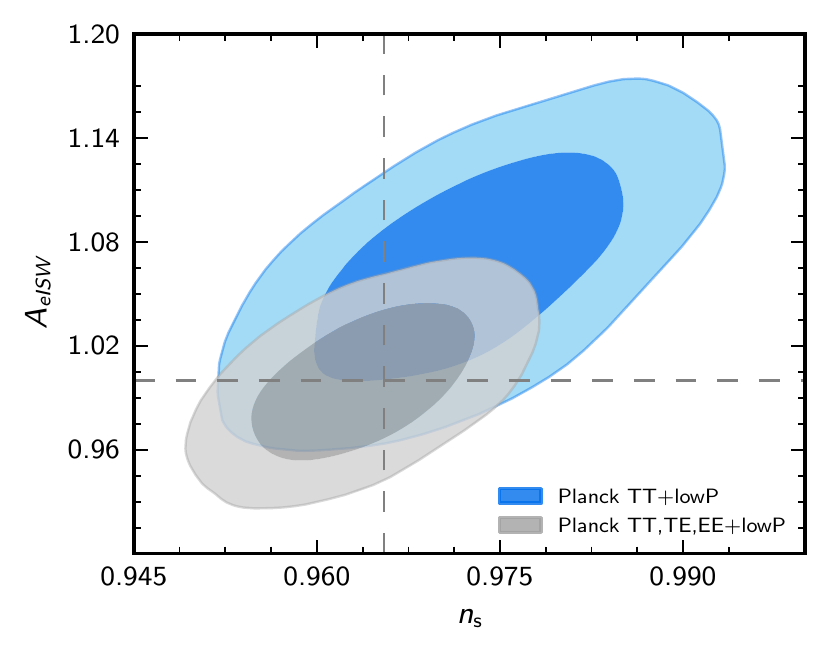}\\
\end{tabular}
\end{center}
\caption{Two-dimensional posterior probability in the ($\Omega_\textup{b} h^2,  A_{e\text{ISW}}$) and ($n_\textup{s}, A_{e\text{ISW}}$) planes for the \textit{Planck} TT+lowP dataset and the \textit{Planck} TT,TE,EE+lowP datasets.}
\label{fig:eisw-2D}
\end{figure*}

By comparing the results given in the first column of Table~\ref{tab:big_tab-2}
with those shown by the Planck Collaboration in \cite{Planck:2015xua} for a $\LCDM$ model, it can be noticed that the most interesting effects which arise from the inclusion of $\aeisw$ as a free parameter are on the parameters $\Omega_\textup{b}h^2$ and $n_\textup{s}$: a lower $\Omega_\textup{b}h^2$ and a higher $n_\textup{s}$ than the standard $\LCDM$ case are favored.

This can be understood looking at Fig.~\ref{fig:eisw-2D}, which shows the correlation between $A_{e\text{ISW}}$ with $\Omega_\textup{b}h^2$ and $n_\textup{s}$. A larger $A_{e\text{ISW}}$ or a larger $\Omega_\textup{b}h^2$ act in (almost) the same way on the CMB spectrum, increasing the height of the peaks at $\ell\sim100$. This is reflected in the strong degeneracy between $A_{e\text{ISW}}$ and $\Omega_\textup{b}h^2$ (left panel of Fig.~\ref{fig:eisw-2D}), in fact a higher value of $A_{e\text{ISW}}$ can be compensated by a decrease of $\Omega_\textup{b}h^2$ to keep fixed the height of the acoustic peaks of the CMB. 
The right panel of Fig.~\ref{fig:eisw-2D} shows the the 68\% CL and 95\% CL allowed regions in the ($n_\textup{s}$, $A_{e\text{ISW}}$) plane: as the value of $n_\textup{s}$ increases, a larger $A_{e\text{ISW}}$ is also allowed.

\begin{figure*}
\begin{center}
\begin{tabular}{c c}
\includegraphics[width=0.98\columnwidth]{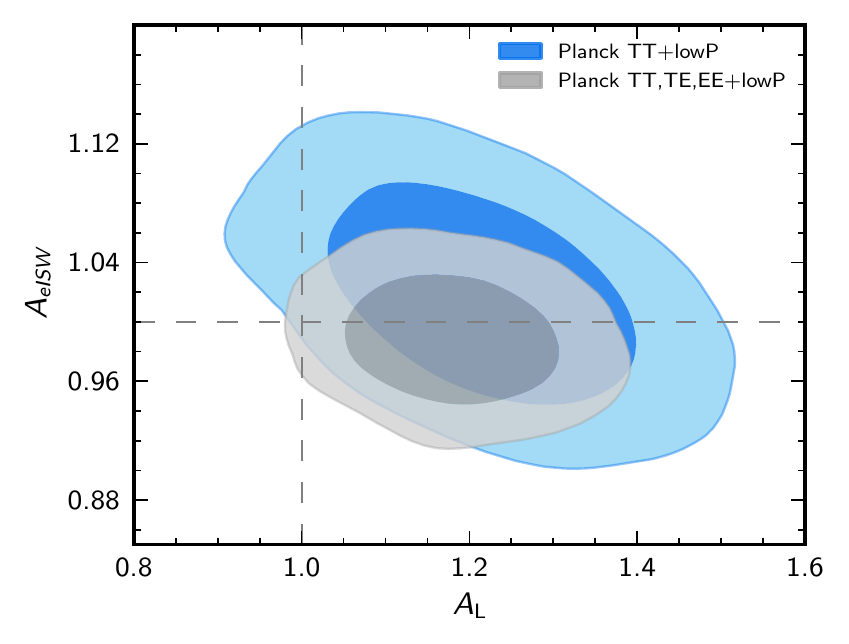}&\includegraphics[width=0.98\columnwidth]{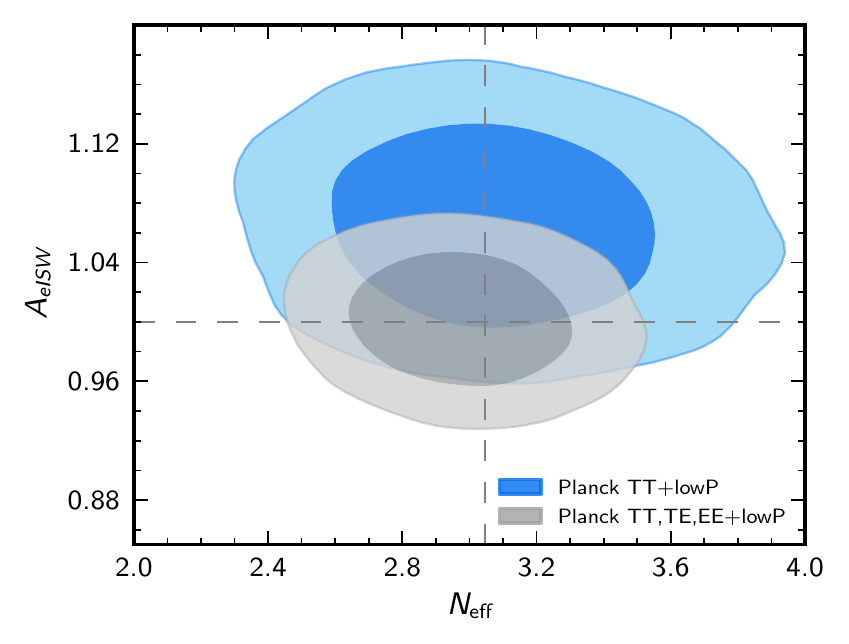}
\end{tabular}
\end{center}
\caption{The left panel depicts the 68\% and 95\% CL allowed regions in the ($A_\textup{L}$, $A_{e\text{ISW}}$) plane for the \textit{Planck} TT+lowP and the \textit{Planck} TT,TE,EE+lowP datasets. The right panel shows the 68\% and 95\% CL regions in the ($N_\textup{eff}$, $A_{e\text{ISW}}$) plane.}
\label{fig:eisw-2D_2}
\end{figure*}

When we consider the  \textit{Planck} TT,TE,EE+lowP datasets (first column of Table~\ref{tab:big_tab-2-TTTEEE}) the bounds on the optical depth, $\tau$, and and the amplitude of the primordial spectrum, $\ln[10^{10} A_s]$, are displaced to higher values and the errors on the cosmological parameters are reduced. 





Instead, as shown in Tables~\ref{tab:big_tab-2} and~\ref{tab:big_tab-2-TTTEEE}, the inclusion of gravitational lensing, $A_\textup{L}$,  and of the effective number of relativistic species, $\neff$, does not change significantly the
constraints on the parameters with respect to those obtained by the Planck Collaboration \cite{Planck:2015xua}. Figure~\ref{fig:eisw-2D_2}, left panel, depicts the 68\%  and 95\% CL allowed
regions in the ($A_\textup{L}$, $A_{e\text{ISW}}$) plane. \textcolor{black}{Even if the early ISW and weak lensing operate at very different scales, the latter is also sensitive to the matter density $\Omega_\textup{m}h^2$ \cite{Ade:2015zua}: this explains the mild correlation between these two parameters shown in the left panel.} The right panel of Figure~\ref{fig:eisw-2D_2} illustrates the 68\% and 95\% CL contours in the ($N_\textup{eff}$, $A_{e\text{ISW}}$) plane resulting from the analysis of CMB data. Notice that,  in contrast to what said in section \ref{subsec:eisw-theory}, these parameters appear uncorrelated. Actually, instead, 
this agrees with the conclusions of \cite{Hou:2011ec}, in which the authors explain how a $Y_P$ ``fixed'' by BBN consistency would not degrade the constraint on $N_\textup{eff}$, even if $A_{e\text{ISW}}$ is left free to vary. 

\begin{figure*}
\begin{center}
\begin{tabular}{c c}
\includegraphics[width=0.98\columnwidth]{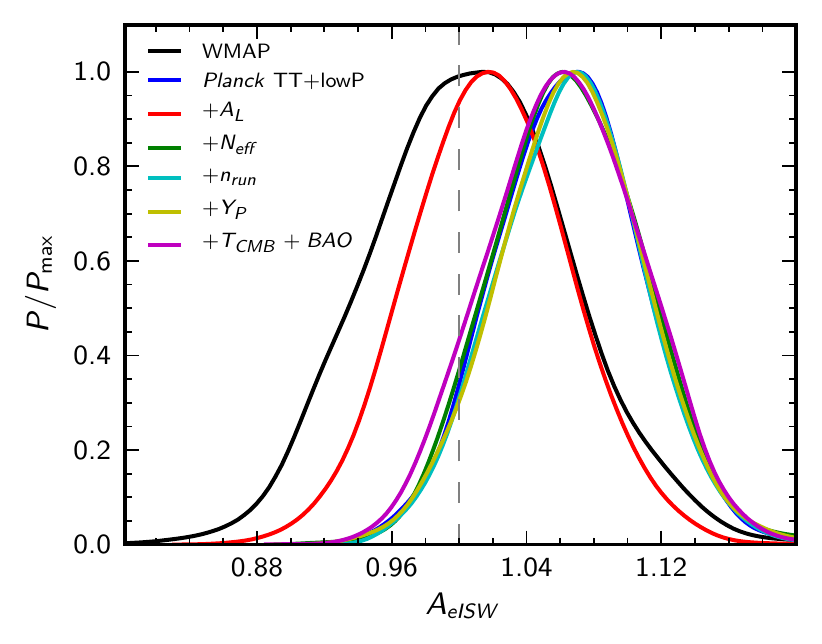}&
\includegraphics[width=0.98\columnwidth]{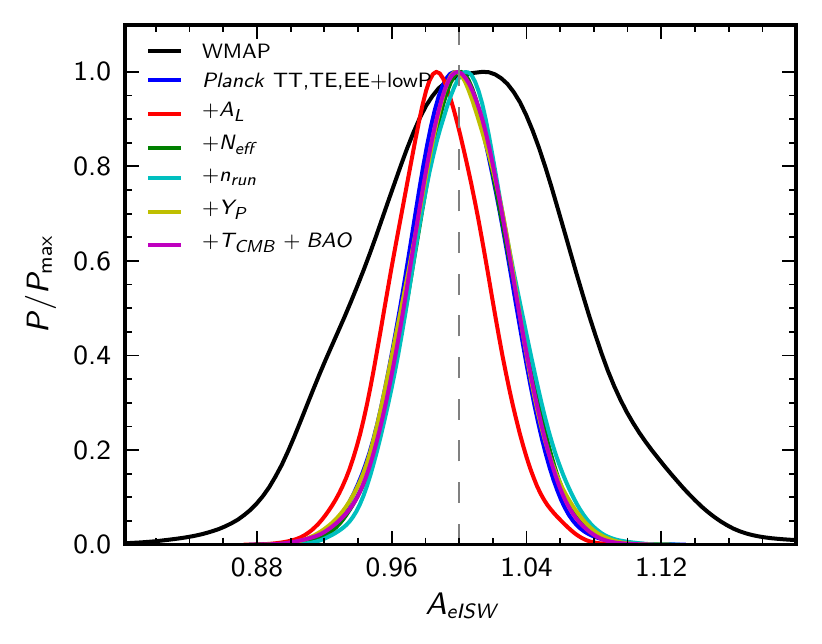}
\end{tabular}
\end{center}
\caption{One-dimensional posterior probability for the amplitude of the early-time ISW effect for the indicated datasets and models. The black and blue curves correspond to a $\Lambda\text{CDM}+A_{e\text{ISW}}$ model. The additional curves come from the indicated one-parameter extension to this baseline model, for the \textit{Planck} TT+lowP dataset (left) and \textit{Planck} TT,TE,EE+lowP dataset (right). When $T_\textup{CMB}$ is varied, BAO datasets \cite{Percival:2009xn, Beutler:2011hx, Blake:2011en, Padmanabhan:2012hf, Anderson:2012sa} are included in the analysis, in order to break degeneracies between cosmological parameters.}
\label{fig:eisw-1D}
\end{figure*}

Tab.~\ref{tab:eisw} depicts the 68\% CL constraints on $A_{e\text{ISW}}$ for the different cosmological models explored in this study using different cosmological data. Firstly, notice that the $\TT$ data alone already provide tighter constraints than WMAP on $A_{e\text{ISW}}$.
Using only the \textit{Planck} TT+lowP data, we can see that the inclusion of the lensing amplitude $A_\textup{L}$ as a free parameter (in addition to the standard $\Lambda\text{CDM}$ picture) tends to diminish the $1\sigma$ indication for a $A_{e\text{ISW}}\neq 1$. 
We also note that this preference persists when we vary other parameters like the effective number of relativistic species $N_\textup{eff}$, the running of the scalar tilt $n_\textup{run}$ and the Helium mass fraction $Y_P$. On the other hand, it vanishes when we consider the  \textit{Planck} TT,TE,EE+lowP data for all different cosmological models. These results are also summarized by the plots of Fig.~\ref{fig:eisw-1D}, showing the one-dimensional posteriors for $\aeisw$ in the various extensions of $\LCDM$ we discussed.

\begin{table}[p]
\begin{center}
\begin{tabular}{c|c}
\hline \hline
Extended Model		&$A_{e\text{ISW}}$\\
$\Lambda$CDM+ & \\
\hline \hline
 $A_{e\text{ISW}}$			& \\
WMAP &                  $1.007^{+0.056}_{-0.058}$\\
\textit{Planck} TT+lowP &$1.064^{+0.042}_{-0.043}$\\
\textit{Planck} TT, TE, EE+lowP  &  $0.999\pm 0.028$ \\
\hline
$A_{e\text{ISW}}+A_\textup{L}$			&\\
\textit{Planck} TT+lowP & $1.018\pm0.046$\\
\textit{Planck} TT, TE, EE+lowP & $0.988\pm0.027$\\
\hline
$A_{e\text{ISW}}+N_\textup{eff}$			& \\
\textit{Planck} TT+lowP  & $1.065\pm0.043$\\
\textit{Planck} TT, TE, EE+lowP & $1.002\pm 0.028$\\
\hline
$A_{e\text{ISW}}+n_\textup{run}$			& \\
\textit{Planck} TT+lowP &$1.066^{+0.041}_{-0.042}$\\
\textit{Planck} TT, TE, EE +lowP & $1.004^{+0.027}_{-0.031}$\\
\hline 
$ A_{e\text{ISW}}+Y_P$			& \\
\textit{Planck} TT+lowP & $1.066\pm0.042$\\
\textit{Planck} TT, TE, EE +lowP &$1.000\pm0.028$\\
\hline
$A_{e\text{ISW}}+T_\textup{CMB}$			&\\
\textit{Planck} TT+lowP+BAO & $1.063\pm0.046$\\
\textit{Planck} TT, TE, EE +lowP+ BAO &$1.001\pm0.028$\\
\hline
\hline
\end{tabular}
\end{center}
\caption{Constraints at 68\% CL on the amplitude of the early-time ISW effect, $A_{e\text{ISW}}$, for the different combinations of datasets and models.}
\label{tab:eisw}
\end{table}



\begin{figure}
\includegraphics[width=0.51\textwidth]{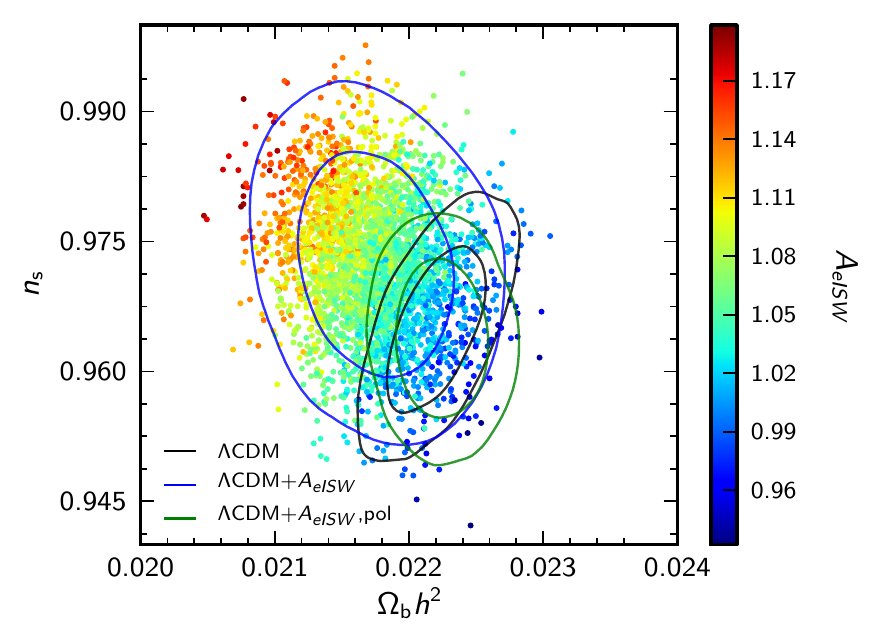}
\caption{\textcolor{black}{Two-dimensional contours in the $\Omega_\textup{b} h^2 - n_\textup{S}$ plane, colored by the value of the parameter $A_{e\text{ISW}}$, for the \textit{Planck} TT+lowP and \textit{Planck} TT,TE,EE+lowP datasets.} The black and blue contours show the two-dimensional posterior probability in the $\Omega_\textup{b} h^2 - n_\textup{S}$ plane for the same dataset and the indicated models. The green contours include the addition of high-$\ell$ polarization.}
\label{fig:eisw-3D}
\end{figure}


Figure~\ref{fig:eisw-3D} shows the 2D marginalized posterior distribution for $\Omega_\textup{b}h^2$ and $n_\textup{s}$ using the \textit{Planck} TT+lowP and \textit{Planck} TT,TE,EE+lowP datasets. We consider two different cosmological models: $\LCDM$ vs. $\LCDM$+$A_{e\text{ISW}}$. Notice that the correlation between $\Omega_\textup{b}h^2$ and $n_\textup{s}$ turns from positive ($\Lambda\text{CDM}$) to negative ($\Lambda\text{CDM} + A_{e\text{ISW}}$). This is due to the strong degeneracy between $\Omega_\textup{b}h^2$ and $A_{e\text{ISW}}$ (already shown in Fig.~\ref{fig:eisw-2D}) that reduces the degeneracies between the other parameters of the $\Lambda\text{CDM}$ model. 
Moreover, if also the information from the Planck high-$\ell$ polarization data is included, the values of these three parameters tend to come in accord with their standard $\LCDM$ value (see Tab.~\ref{tab:eisw}), even if the direction of the degeneracy between $\Omega_\textup{b}h^2$ and $n_\textup{s}$ remains positive.


\subsection{Late ISW -- results} \label{subsec:lisw-results}

\begin{figure*}
\begin{center}
\begin{tabular}{c c}
\includegraphics[width=0.95\columnwidth]{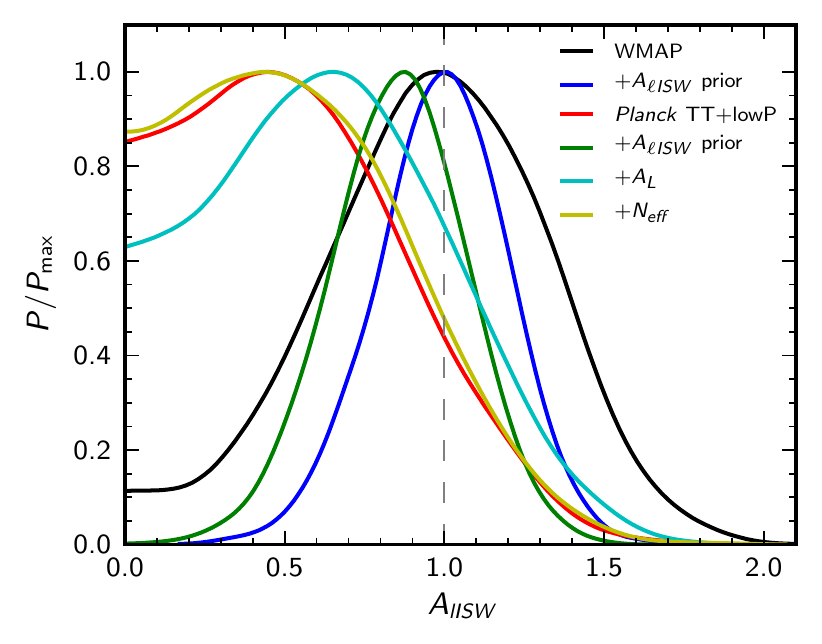}&
\includegraphics[width=0.95\columnwidth]{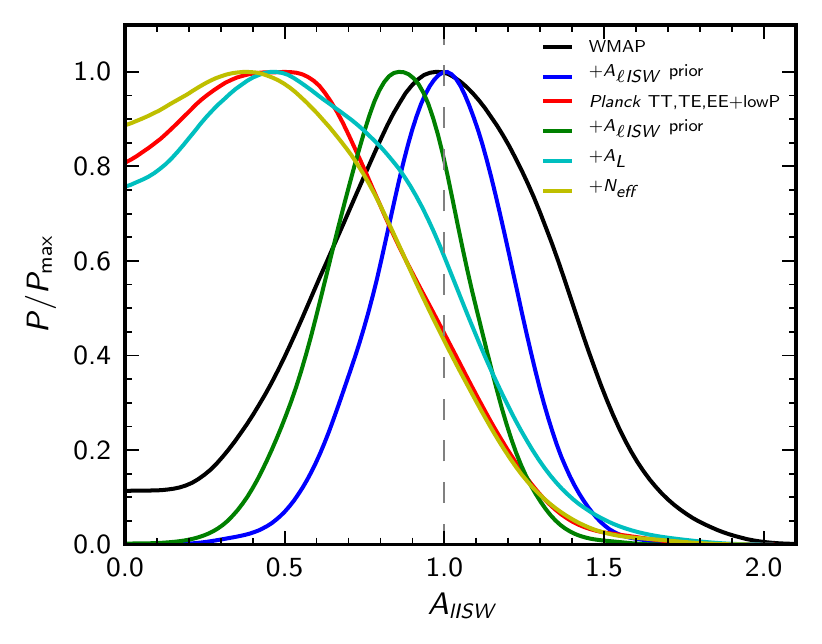}\\
\end{tabular}
\end{center}
\caption{One-dimensional posterior probability for the amplitude of the late-time ISW effect for the data sets and models discussed in the text. The black curves refer to a $\Lambda\text{CDM}+A_{l\text{ISW}}$, with the WMAP data set for the high-$\ell$ polarization. The remaining curves include the \textit{Planck}TT+lowP data (left panel), and the \textit{Planck} TT,TE,EE+lowP data (right panel). The ``$A_{l\text{ISW}}$ prior'' label indicates the inclusion of the gaussian prior on $A_{l\text{ISW}}$ coming from the cross-correlated analysis of the CMB bispectrum and galaxy clusters.}
\label{fig:lisw-1D}
\end{figure*}

In this section we present the results obtained considering only the late ISW effect. Table~\ref{tab:lisw-TT+TTTEEE} presents the constraints on $A_{l\text{ISW}}$ for the different cosmological data combinations considered here. Fig.~\ref
{fig:lisw-1D} contains the one-dimensional posteriors for the amplitude of the late-time ISW effect in the various extensions of $\Lambda$CDM model. Notice that when we consider the case with a flat prior on $\alisw$, there is consistency with $A_{l\text{ISW}} = 1$ for the WMAP dataset. The \textit{Planck} TT+lowP and the \textit{Planck} TT, TE, EE+lowP measurements set  the 95\% CL upper limit of  $A_{l\text{ISW}}\lesssim 1.14$  and $A_{l\text{ISW}}\lesssim 1.11$ respectively. We note that Planck alone does not improve significantly the constraint on $A_{l\text{ISW}}$ with respect to WMAP measurements. This occurs because the late-time ISW affects a region of CMB power spectrum multipoles that is dominated by cosmic variance, rather than by instrumental precision. Moreover the bounds on $A_{l\text{ISW}}$ are not affected if the effective number of relativistic species ($N_\textup{eff}$) is included.

We also consider a gaussian prior of $A_{l\text{ISW}} = 1.00\pm0.25$ from the bispectrum-LSS cross-correlation analysis, which allows us to take into account the constraints on the late ISW coming from large-scale structure measurements. The inclusion of the prior results in a tighter constraints from WMAP, while the posterior on $A_{l\text{ISW}}$ when Planck data set is considered is shifted towards $A_{l\text{ISW}}=1$.

\begin{table}[h]
\begin{center}
\begin{tabular}{c|c}
\hline \hline
Extended Model		&$A_{l\text{ISW}}$\\
$\Lambda$CDM+ & \\
\hline \hline
 $A_{l\text{ISW}}$			& \\
WMAP &               $0.958^{+0.391}_{-0.317}$  \\
\textit{Planck} TT+lowP   &$<1.14$\,(95\%\,CL)\\
\textit{Planck} TT, TE, EE+lowP  &  $<1.11$\,(95\%\,CL) \\
\hline
$A_{l\text{ISW}}$, prior			&\\
WMAP &   $0.958^{+0.220}_{-0.192}$\\       
\textit{Planck} TT+lowP & $0.853\pm0.211$\\
\textit{Planck} TT, TE, EE+lowP &$0.847_{-0.203}^{+0.217}$\\
\hline
$A_{l\text{ISW}}+N_\textup{eff}$			& \\
\textit{Planck} TT+lowP  &$<1.14$\,(95\%\,CL)\\
\textit{Planck} TT, TE, EE+lowP  &$<1.11$\,(95\%\,CL)\\
\hline
$A_{l\text{ISW}}+A_\textup{L}$			& \\
\textit{Planck} TT+lowP &$<1.25$\,(95\%\,CL)\\
\textit{Planck} TT, TE, EE +lowP &$<1.12$\,(95\%\,CL)\\
\hline
\hline
\end{tabular}
\end{center}
\caption{Constraints at 68\% CL (unless otherwise stated) on the amplitude of the  late-time ISW effect, $A_{l\text{ISW}}$, for the different combinations of data sets and models considered in the text.}
\label{tab:lisw-TT+TTTEEE}
\end{table}

\begin{figure}
\includegraphics[width=0.42\textwidth]{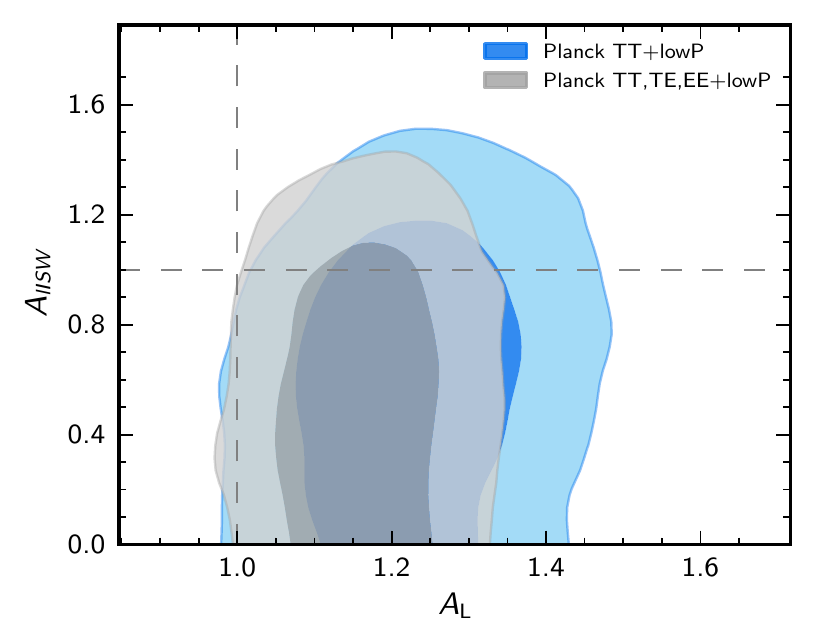}
\caption{Two-dimensional posterior probability in the $\alens-\alisw$ plane for the \textit{Planck} TT+lowP and \textit{Planck} TT,TE,EE+lowP datasets. This posterior shows that, while the amplitude of the late ISW effect and the lensing parameter $A_\textup{L}$ are not correlated, the inclusion of high-$\ell$ polarization data from Planck brings the contours back in accord with $\alens = 1$ and $\alisw = 1$.}
\label{fig:lisw-2D}
\end{figure}

Fig.~\ref{fig:lisw-2D} shows the 68\% and 95\%~CL allowed regions in the ($A_\textup{L}, A_{l\text{ISW}}$) plane for the \textit{Planck} TT+lowP and \textit{Planck} TT,TE,EE+lowP data sets. Notice that there is no correlation between $A_\textup{L} - A_{l\text{ISW}}$. This was expected since the late ISW is active at low $\ell$, while weak lensing operates at high $\ell$. Moreover there is a mild preference for a non-standard value 
of both parameters. Marginalizing over $A_\textup{L}$ we obtain an upper limit of $A_{l\text{ISW}} < 1.25$ at 95\%\,CL using the \textit{Planck} TT+lowP data set, while the inclusion of high-$\ell$ polarization measurements tightens the constraint at $A_{l\text{ISW}} < 1.12$ at 95\%~CL.



\subsection{Early $+$ late ISW} \label{subsec:eisw+lisw}

\noindent 

\begin{figure*}
\begin{center}
\begin{tabular}{c c} 
\includegraphics[width=0.95\columnwidth]{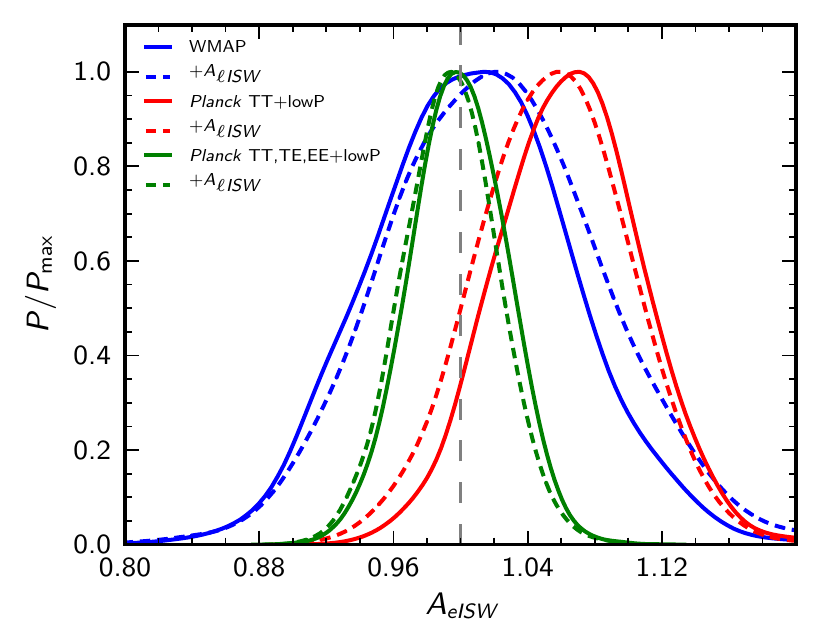}&
\includegraphics[width=0.95\columnwidth]{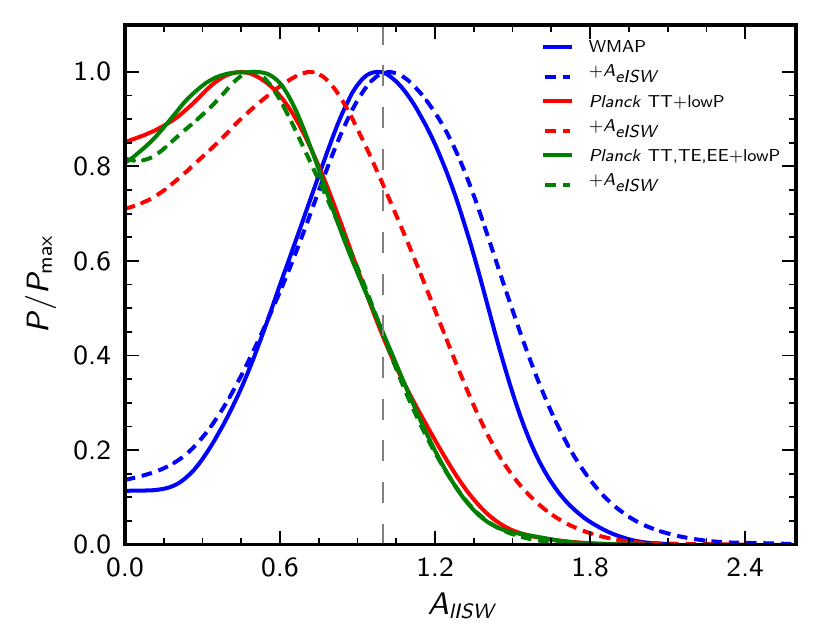} \\
\end{tabular}
\end{center}
\caption{One-dimensional posterior probability for the amplitude of the early-time ISW effect (left) and late-time ISW effect (right) for the indicated datasets and models. For each plot, the solid curves have been obtained for the corresponding one-parameter extension to the base $\Lambda$CDM model. The dashed curves correspond to the joint variation of $A_{e\text{ISW}}$ and $A_{l\text{ISW}}$.}
\label{fig:eisw+lisw-1D_1_e_2D}
\end{figure*}

We conclude by considering the case of both $A_{e\text{ISW}}$ and $A_{l\text{ISW}}$ varying jointly. Constraints on these two parameters are reported in Tab.~\ref{tab:eisw+lisw}. The one-dimensional and two-dimensional posterior probabilities for a selected subset of datasets and models are shown in Fig.~\ref{fig:eisw+lisw-1D_1_e_2D}. As mentioned in Sec.~\ref{subsec:eisw-results} and ~\ref{subsec:lisw-results}, when compared with the results from WMAP, the Planck data provide much tighter constraints on $A_{e\text{ISW}}$ even when considering temperature only, while the constraining power on $A_{l\text{ISW}}$ is comparable.

The upper bounds on $A_{l\text{ISW}}$ are well compatible with the standard case for all datasets used, while there is a $1\sigma$ preference of $A_{e\text{ISW}}\neq 1$ when using the $\TT$ dataset. We note, though, that such a preference for $A_{e\text{ISW}}\neq 1$ disappears when we let $A_\textup{L}$ free to vary, as a result of the mild degeneracy between the two parameters discussed in Sec.~\ref{subsec:eisw-results}. Allowing the number of relativistic species to vary does not alter the constraints with respect to the minimal extension to $\Lambda$CDM. 

\begin{table*}
\caption{Constraints at 68\% CL (unless otherwise stated) on the amplitude of the late-time ISW effect $A_{l\text{ISW}}$ and of the early-time ISW effect $A_{e\text{ISW}}$ for the indicated datasets and models.} 
\begin{center}
\begin{tabular}{c|c|c}
\hline \hline
Dataset, model		&$A_{l\text{ISW}}$		&$A_{e\text{ISW}}$\\
\hline \hline
WMAP, $\Lambda\text{CDM} + A_{l\text{ISW}}+A_{e\text{ISW}}$			&$1.011^{+0.434}_{-0.374}$			&$1.019^{+0.061}_{-0.066}$\\
Planck $\TT$, $\Lambda\text{CDM} + A_{l\text{ISW}}+A_{e\text{ISW}}$			&$<1.34\,(95\%\,CL)$			&$1.055\pm0.044$\\
Planck $\TT$, $\Lambda\text{CDM}+A_{l\text{ISW}} + A_{e\text{ISW}} + A_\textup{L}$			&$<1.32\,(95\%\,CL)$			&$1.009^{+0.047}_{-0.048}$\\
Planck $\TT$, $\Lambda\text{CDM}+A_{l\text{ISW}} + A_{e\text{ISW}} + N_\textup{eff}$			&$<1.35\,(95\%\,CL)$			&$1.057^{+0.043}_{-0.044}$\\
Planck $\TTTEEE$, $\Lambda\text{CDM} + A_{l\text{ISW}}+A_{e\text{ISW}}$			&$<1.11\,(95\%\,CL)$			&$0.994^{+0.027}_{-0.028}$\\
Planck $\TTTEEE$, $\Lambda\text{CDM} + A_{l\text{ISW}} + A_{e\text{ISW}} + A_\textup{L}$			&$<1.12\,(95\%\,CL)$			&$0.985\pm 0.028$\\
Planck $\TTTEEE$, $\Lambda\text{CDM} + A_{l\text{ISW}} + A_{e\text{ISW}} + N_\textup{eff}$			&$<1.10\,(95\%\,CL)$			&$0.996_{-0.030}^{+0.028}$\\
\hline
\hline
\end{tabular}
\end{center}
\label{tab:eisw+lisw}
\end{table*}

\begin{figure}
\includegraphics[width=0.43\textwidth]{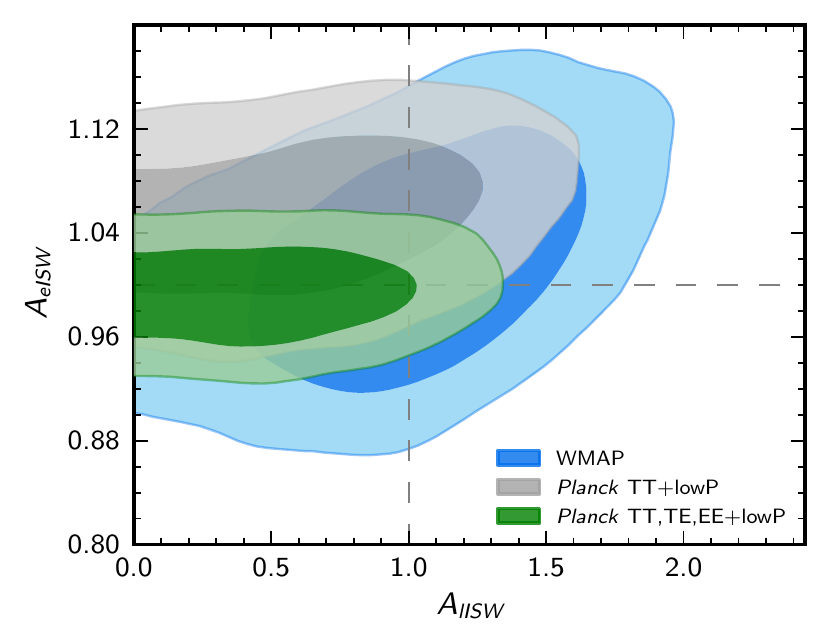}
\caption{Two-dimensional posterior probability in the $A_{l\text{ISW}} - A_{e\text{ISW}}$ plane for the \textit{Planck} TT+lowP and \textit{Planck} TT,TE,EE+lowP datasets.}
\label{fig:lisw+eisw-2D}
\end{figure}
The inclusion of small-scale polarization data significantly tightens the constraints on $A_{e\text{ISW}}$, almost halving the posterior width. On the other hand, as already expected, it does not provide further information on $A_{l\text{ISW}}$, as highlighted by the superposition of both the green curves with the solid red one in the top right panel of Fig.~\ref{fig:eisw+lisw-1D_1_e_2D}.

%

\section{Conclusions} \label{sec:concl}

\noindent In this paper we study the constraints on the amplitude of the Integrated Sachs Wolfe effect, both its early and late time contributions.

We find that the $\TT$ data is consistent with a non-zero early ISW, with an amplitude $A_{e\text{ISW}}$ in agreement with $A_{e\text{ISW}} = 1$ as predicted by theory, with a $1\sigma$ preference of $A_{e\text{ISW}}\neq 1$ when considering extensions to the $\LCDM$ model discussed in this work. We also confirm the strong degeneracy between the amplitude of the early ISW and parameters like $\Omega_\textup{b} h^2$ and $n_\textup{S}$. Our analysis also hints for a correlation between $A_{e\text{ISW}}$ and the lensing parameter $A_\textup{L}$. 

Regarding the late ISW, Planck data alone place a constraint $A_{l\text{ISW}}\lesssim 1.1$ at 95\% CL. When supplemented with a prior on $A_{l\text{ISW}}$ coming from CMB temperature anisotropies-weak lensing correlations, however, we find a $\sim 4\sigma$ detection $A_{l\text{ISW}} = 0.85\pm0.21$.

When we consider also the recent polarization data at high $\ell$ from the Planck collaboration, we find that the evidences for a non-standard value of $A_{e\text{ISW}}$ disappear. The reason is that the addition of $TE$ and $EE$ spectra leads to a better agreement of data with the standard $\Lambda$CDM model. More precisely, $A_{e\text{ISW}}$ gets dragged towards $1$ through its degeneracy with $\Omega_\tu{b}h^2$ and $n_\textup{s}$, which return in agreement with the $\Lambda$CDM best fit when polarization is included.

On the other hand, using the small-scale polarization spectra does not change the results obtained for $A_{l\text{ISW}}$. Their effect is to slightly tighten the upper bounds obtained when considering only the temperature spectra.

When the two parameters are allowed to vary jointly, the same pattern described above is reproduced.

\subsection*{Acknowledgements} \label{subsec:acknow}

\noindent We would like to thank Antony Lewis for the use of the numerical codes \texttt{cosmomc} and \texttt{camb}. We acknowledge support by the research grant Theoretical Astroparticle Physics number 2012CPPYP7 under the program PRIN 2012 funded by MIUR and by TASP, iniziativa specifica INFN.



\begin{thebibliography}{99}
\bibitem{footnote:1}
 \textcolor{black}{We use the Newtonian gauge convention $g_{00} = -1 - 2\Psi$ and $g_{ij} = a^2\delta_{ij}(1+2\Phi)$.}
\bibitem{footnote:2}
  At the present time the relevant length of photon trajectories is comparable to the horizon size, so that the main contribution to Eq.~\eqref{eq:1} comes from perturbations with $\lambda\sim\eta_0$ (the contributions of shorter wavelengths again average away). Therefore the late ISW effect influences only multipoles corresponding to $k\sim 1/\eta_0$, that is $\ell\sim 1$.
\end{thebibliography}
\end{document}